\documentclass[10pt]{iopart}

\expandafter\let\csname equation*\endcsname\relax      % These 2 lines prevent equation errors from
\expandafter\let\csname endequation*\endcsname\relax   % clash of IOPART package & AMS packages
\usepackage{iopams}
\usepackage[utf8]{inputenc}
\usepackage[english]{babel}
\usepackage[T1]{fontenc}
\usepackage{amsmath}
\usepackage{amsfonts}
\usepackage{amssymb}
\usepackage{graphicx}  
\usepackage{subcaption} 
\usepackage{tensor}
\usepackage{tikz}
\usetikzlibrary{decorations.pathmorphing}
\usepackage{hyperref}

\begin{document}

\title[Phantom scalar field counterpart to Curzon--Chazy spacetime]{Phantom scalar field counterpart to Curzon--Chazy spacetime}

\author{Luk\'{a}\v{s} Polcar and Otakar Sv\'{\i}tek}
\address{Institute of Theoretical Physics, Faculty of Mathematics and Physics, Charles University, V Hole\v{s}ovi\v{c}k\'{a}ch 2, 180 00 Prague 8, Czech Republic}
\eads{\mailto{polcar.vm@seznam.cz}, \mailto{ota@matfyz.cz}}

%%%%%%%%%%%%%%%%%%%%%%%%%%%%%%%%%%%%%%%%%%%%%%
%\author{Luk\'{a}\v{s} Polcar\footnote{polcar.vm@seznam.cz}} 
%\author{Otakar Sv\'{\i}tek \footnote{ota@matfyz.cz}} 

%\affil{Institute of Theoretical Physics, Faculty of Mathematics and Physics, Charles University, V Hole\v{s}ovi\v{c}k\'{a}ch 2, 180 00 Prague 8, Czech Republic}

%\begin{document}

%\maketitle
%%%%%%%%%%%%%%%%%%%%%%%%%%%%%%%%%%%%%%%%%%%%%

\begin{abstract}
    We derive and analyze phantom scalar field counterpart to Curzon--Chazy spacetime. Such solution contains a wormhole throat while the region inside the throat behaves like a one-directional time machine. We describe its conformal structure and non-scalar singularity hidden inside the wormhole. We examine the results provided by different definitions of mass of the spacetime to understand their value in the presence of phantom matter. {The electromagnetic generalization of this spacetime is as well briefly considered.}
\end{abstract}

\section{Introduction}
Scalar field provides a canonical source for spacetime since it is both sufficiently simple and reasonably realistic source. The simplicity enables one to employ analytical methods more thoroughly and also presents much cleaner picture of potential physical effects. At the same time there is both experimental verification of the role of scalar fields in nature (Higgs field) and an extensive theoretical use in modeling of various phenomena (inflation, dark energy/matter).

Substantial interest has been naturally devoted to spherically symmetric solutions of gravity coupled to scalar field with the no-hair theorems being one of the main focus (e.g. see recent review \cite{Herdeiro-review}). Obviously, one would like to analytically investigate solutions with less symmetry. Natural candidate in this regard is the Weyl family of metrics and its generalizations including scalar field (both standard and phantom) which was recently studied in \cite{Gibbons:2017jzk}. One aim of this analysis was to uncover the role these geometries might play in building wormholes. Even more ambitious generalizations of spherically symmetric solutions with scalar fields have been performed recently (e.g. towards dynamical spacetimes without symmetries \cite{Tahamtan-PRD:2015,Tahamtan-PRD:2016}) however in these generic situations explicit analysis of basic properties becomes extremely complicated (e.g. even the proof of the existence of horizon demands lengthy mathematical analysis).

The simplest solution of the Weyl class \cite{Weyl} is from a certain point of view the Curzon--Chazy spacetime \cite{Curzon} since it produces Newtonian potential. On the other hand it contains directional singularity at the origin. Naturally, one would like to obtain similarly simple scalar field solution. After recalling the general solution for Curzon--Chazy geometry with minimally coupled massless scalar field we investigate a specific limit (made possible in the presence of phantom scalar field) that reduces this two-parametric solution to one-parametric. In this solution the Newtonian potential is preserved and is apparently sourced by the scalar field which has Newtonian character as well. In this solution the singularity at the origin is removed and moreover the spacetime becomes spherically symmetric.

At the same time, this solution represents certain limit of the famous Janis--Newman--Winicour (JNW) solution \cite{JNW} (when generalized to phantom scalar field) which was however initially discovered by Fisher \cite{Fisher}. JNW solution with standard scalar field serves as a prime example of how the scalar field spoils regularity of a horizon or leads to naked singularities. This behavior of scalar field is hard to tame even when invoking other sources (see \cite{Tahamtan-PRD:2020}).

In this paper, we derive the phantom scalar field version of Curzon--Chazy spacetime and proceed to study its features. Namely we show that it contains a wormhole throat and works as a one-directional time machine. We analyze its conformal structure, discover non-scalar singularity hidden on the other side of wormhole throat and devote substantial effort to understand the value of different definitions for energy of the spacetime and analyze the interplay of scalar field and gravitational parts of the total energy. {Note, that the wormhole we are having in mind is of the bridge type (with general prototype being discussed at the end of \cite{Tahamtan-PRD:2016}) as opposed to the thin-shell type (e.g. \cite{Tahamtan-EPJC:2018}).} Finally, we use a recently published study \cite{Maeda} about generating techniques to obtain electromagnetic generalization of the solution and analyze its drastically modified conformal structure. 

 {We do realize that having a solution with phantom  scalar field is hard to justify physically. In particular the system has energy unbounded from below which would lead to instability, see e.g. \cite{Cline:2003} for the discussion of serious quantum instabilities affecting such matter systems. This work was motivated purely theoretically, that is, by a particular limit of the well-known Curzon-Chazy solution generalized to contain scalar field. However, we briefly analyze some astrophysically relevant properties, e.g. gravitational lensing.}

\section{Obtaining the solution}

We consider a massless scalar field minimally coupled to gravity described by action

\begin{equation}\label{action}
S=\displaystyle\int(\mathcal{R}-2\varepsilon \nabla_\mu\Phi \nabla^\mu\Phi)\sqrt{-g}\, \mathrm{d^4}x.
\end{equation}  
where the parameter $\varepsilon$ enables us to describe either the conventional scalar field ($\varepsilon=1$) or the so called phantom scalar field ($\varepsilon=-1$) which has negative energy density. 
 {The scalar field $\Phi$ is related to its usual definition   $\Psi$ by $\Phi=\sqrt{\frac{\kappa}{2}} \Psi$ where the Einstein gravitational constant is $\kappa=8\pi$, with $G=1$ and $c=1$.}
The equations of motion then take form

 \begin{equation}\label{Einsteineq}
R_{\mu \nu}=2\varepsilon\partial_{\mu}\Phi \partial_{\nu}\Phi,\hspace{15pt} \square \Phi=0.
\end{equation} 

We will now restrict ourselves to a static axially symmetric spacetime with a line element in the form introduced by Weyl \cite{Weyl}

\begin{equation}\label{Weylmetric}
\displaystyle\mathrm{d}s^2=-e^{2U(\rho,z)}\mathrm{d}t^2+e^{2k(\rho,z)-2U(\rho,z)}(\mathrm{d}\rho^2+\mathrm{d}z^2)+\rho^2 e^{-2U(\rho,z)} \mathrm{d}\phi^2.
\end{equation}  
This metric is determined by only two functions $U(\rho,z)$ and $k(\rho,z)$.

The field equations \eqref{Einsteineq} are then reduced to \cite{Gibbons:2017jzk}

\begin{subequations}\label{Weyleq}
\begin{gather}
\label{Weyl1}
 U_{,\rho\rho}+\frac{1}{\rho}U_{,\rho}+U_{,zz}=0,\hspace{15pt} \Phi_ {,\rho\rho}+\frac{1}{\rho}\Phi_{,\rho}+\Phi_{,zz}=0 \\
 \label{Weyl2}
  k_{,\rho}=\rho\lbrace \big[(U_{,\rho})^2-(U_{,z})^2\big]+ \varepsilon\big[(\Phi_{,\rho})^2-(\Phi_{,z})^2\big]\rbrace,\hspace{15pt} k_{,z}=2\rho( U_{,\rho}U_{,z}+\varepsilon\Phi_{,\rho}\Phi_{,z}).
 \end{gather}
\end{subequations}

Since equations \eqref{Weyl1} are Laplace equations in flat 3-dimensional space expressed for axially symmetric situation in polar coordinates, the first solution that may come to one's mind is that of a Newtonian gravitational potential of a point particle located at the origin. For the vacuum case $\Phi=0$
we can solve the corresponding $k(\rho,z)$ using \eqref{Weyl2} arriving at

\begin{equation}\label{Curzon}
 U=-\frac{M}{r}, \hspace{15pt}  k=-\frac{M^2\rho^2}{2r^4}, \hspace{15pt} r=\sqrt{\rho^2+z^2}.
\end{equation} 

This is the (single-particle) Curzon--Chazy  solution \cite{Curzon}. The most striking feature of this solution  is the direction-dependent singularity at $r=0$. When one approaches the origin from the equatorial plane ($z=0, \rho \rightarrow 0^+$) the Kretschmann scalar diverges. On the other hand if one approaches the origin along the z-axis using an appropriate worldline the singularity can be avoided and the spacetime can be extended beyond $r=0$. The structure of the singularity was studied  in \cite{Scott:1985ad} and more recently in \cite{Taylor:2005fz}.\par
We can now add the scalar field and see how it affects the singularity. The equations \eqref{Weyl2} possess a rotation ($\varepsilon=1$) or a boost symmetry ($\varepsilon=-1$) for the pair $(U, \Phi)$ while preserving the function $k$ (details  in \cite{Gibbons:2017jzk}). This together with the fact that $\Phi$ satisfies the same equation as $U$ allows us to generate a solution with scalar field from the vacuum solution \eqref{Curzon}. A solution with scalar field is then 

\begin{equation}\label{Curzon+scalar}
 U=-\frac{1}{s}\frac{M}{r}, \hspace{15pt}  k=-\frac{M^2\rho^2}{2r^4}, \hspace{15pt} \Phi =\pm\frac{\sqrt{\varepsilon(s^2-1})}{s}\frac{M}{r}
\end{equation} 

where the new parameter $s$ satisfies $\vert s \vert >1$ for $\varepsilon=1$ while for the phantom scalar field we have  $\vert s \vert <1$. 

The solution \eqref{Curzon+scalar} has two interesting limits. The first is the ultrastatic limit $\vert s \vert \rightarrow \infty$ which eliminates the function $U$ completely. The Kretschmann scalar is then 

\begin{equation}\label{Kretschmann_ultrastat}
K^{ustat}=\,{\frac {12{M}^{4}}{{r}^{8}}{{\rm e}^{\,{\frac {2{M}^{2}{\rho}^{2}}{{
r}^{4}}}}}}.
\end{equation} 
It is clear that $K^{ustat}\rightarrow \infty$ for $r \rightarrow 0$ and the corresponding singularity is thus no longer direction-dependent.\par
The other limit (described in \cite{Gibbons:2017jzk}) eliminates the metric function $k$. This can be done by reparametrizing $M$ and evaluating the following limit

\begin{equation}\label{limit}
M=m s \quad,\quad s \rightarrow 0\ ,
\end{equation} 
while keeping $m>0$ fixed.
The Kretschmann scalar now reads

\begin{equation}\label{Kretschmann}
K^{scc}=\,{\frac {{{4\rm e}^{\,{\frac {-4m}{r}}}}{m}^{2} \left( 7\,{m}^{2}-16\,mr+12\,{r}^{2} \right) }{{r}^{8}}}.
\end{equation} 

We have thus managed to seemingly remove the curvature singularity at $r=0$ where $K^{scc}$ is zero in the sense of the corresponding limit. In addition, the spacetime is now spherically symmetric. Using the transformation to spherical coordinates in much the same way as in the Euclidean space we arrive at the Spherical Curzon--Chazy spacetime (SCC)

\begin{equation}\label{scc}
\displaystyle\mathrm{d}s^2=-e^{-\frac{2m}{r}}\mathrm{d}t^2+e^{\frac{2m}{r}}(\mathrm{d}r^2+r^2 \mathrm{d}\Omega^2), \hspace{15pt} \Phi=\pm\frac{m}{r}.
\end{equation}  

Although the initial idea was to figure out how the presence of scalar field affects the singularity, the SCC can be also obtained from the Janis--Newman--Winicour (JNW) solution

\begin{subequations}\label{JNW}
\begin{gather}
\displaystyle\mathrm{d}s^2=-\left( 1-\,{\frac {2M}{r}} \right) ^{{\frac {1}{s}}}\mathrm{d}t^2+\left( 1-\,{\frac {2M}{r}} \right) ^{{-\frac {1}{s}}}\mathrm{d}r^2+\left( 1-\,{\frac {2M}{r}} \right) ^{{-\frac {1}{s}}}(r^2-2Mr)\mathrm{d}\Omega^2, \hspace{15pt}  \\ \Phi=\pm \frac{1}{2}\,{\frac {\sqrt {\vert 1-{s}^{2}\vert}}{s}\ln  \left( 1-\,{\frac {2M}{r}}
 \right) }.
\end{gather}
\end{subequations}

The JNW solution  \cite{JNW} is a famous static spherically symmetric solution with a free scalar field which was initially discovered by Fisher \cite{Fisher} but later rediscovered several times. The intervals of $s$ corresponding to the conventional or the phantom scalar field are the same as above. If we perform the limit \eqref{limit} we again recover the SCC solution \eqref{scc}. 
\par
There is yet another class of solutions to which the SCC spacetime belongs. This solution was found by Gibbons in \cite{ Gibbons:2003yj} where the corresponding metric reads

 \begin{equation}\label{Gibbons}
\displaystyle\mathrm{d}s^2=-e^{2 U(X)}\mathrm{d}t^2+e^{-2 U(X)}\displaystyle\mathrm{d}l^2, \hspace{15pt} \Phi=\pm U(X).
\end{equation}  
The line element $\displaystyle\mathrm{d}l^2$ is that of flat space while the function $U(X)$ satisfies ordinary flat-space Laplace equation $\triangle U(X)=0$. We can thus have multi-particle solution where the gravitational attraction is compensated by the repulsion due to the presence of the phantom scalar field. 
This class of solution is analogical to the Majumdar–Papapetrou class  \cite{Majumdar}  where the extremely charged particles mutual gravitational attraction is balanced by their electric repulsion. In this analogy the Spherical Curzon--Chazy spacetime corresponds to the extreme Reissner--Nordstr\"{o}m   black hole.\par

\section{Basic properties and geodesics}
The SCC metric asymptotically behaves like the Schwarzschild solution. This becomes clear when using the areal radius defined below by expression \eqref{radius} as a coordinate instead of $r$. Naturally, in the limit $r\rightarrow \infty$ the solution reduces to the Minkowski metric. The key point of our investigation thus lies on the other side, at $r=0$ where (at the least) a coordinate singularity is present. The limits $r \rightarrow 0$ of $g_{tt}$ and $g_{rr}$ are the same as in the case of the Schwarzschild horizon where the singularity can be eliminated when using suitable coordinates.  Since the Kretschmann scalar \eqref{Kretschmann} is zero at $r=0$ one would naively expect the same can be done for the SCC spacetime. \par
On the other hand, not everything is similar to the Schwarzschild metric. One obvious difference that one may notice is the divergence of the areal radius of the sphere centered at $r=0$

 \begin{equation}\label{radius}
R_A(r)=r e^{\frac{m}{r}}.
\end{equation}  
The radius reaches its minimum at $r=m$ and then grows again to infinity. This behaviour is often typical for wormholes and it will be discussed in the following sections.\par
We shall now try to reach $r=0$ using various radial geodesics. Starting from $r=r_0>0$ the proper distance to the origin along a spacelike geodesic given by $t=const.$ is 
\begin{equation}\label{properdistance}
d(0,r_0)=\displaystyle\int\limits_{0}^{r_0} e^{\frac{m}{r}} \mathrm{d}r =+\infty.
\end{equation} 
This is in direct contrast  to Schwarzschild where the distance to the horizon is finite. For an ingoing null radial geodesic parametrised by an affine parameter $\lambda$ the tangent vector  reads 

\begin{equation}\label{lmu}
l^\mu =\frac{\mathrm{d}x^\mu}{\mathrm{d}\lambda}= (e^{\frac{2m}{r}},-1,0,0),
\end{equation}
which shows that $r=0$ can only be reached in infinite coordinate time $t$. On the other hand light arrives at $r=0$ in a finite value of $\lambda$. The same conclusions apply also for the timelike geodesics with four-velocity in the form

\begin{equation}\label{fourvelocity}
U^\mu=(E e^{\frac{2m}{r}},\pm\sqrt{E^2-e^{\frac{-2m}{r}}},0,0)
\end{equation} 
where $E$ is the energy with respect to the timelike Killing field $\xi^{(t)}_\mu$

 $$E=-U^\mu\xi^{(t)}_\mu = -g_{tt}  U^t.$$
The proper time to reach $r=0$ is then finite 
 
 \begin{equation}\label{propertime}
\tau=\int_{r_0}^{0} -\frac{\mathrm{d}r}{\sqrt{E^2-e^{\frac{-2m}{r}}}} < +\infty.
\end{equation}

This creates a counter-intuitive notion that an infinite distance can be covered in a finite proper time. It also points to the geodesic incompleteness of the spacetime.\par
As was already mentioned above the finiteness of the Kretschmann scalar  \eqref{Kretschmann}   at $r=0$ could indicate that the coordinate singularity can be removed. For the radial part of the metric it is indeed so and for that we can use the proper time $\tau$ of timelike radial geodesics as one coordinate while for the other coordinate $l$ we find  the corresponding tangent vector $L^\mu$ satisfying {(we use $[\cdot ,\cdot]$ to denote a commutator of two vector fields)}

 \begin{equation}\label{commutation}
 [U,L]=0, \hspace{15pt} U^\mu L_\mu=0.
\end{equation}

The simplest form of $L^\mu$ is then
 \begin{equation}\label{Lmu}
L^\mu=(\pm (E e^{\frac{2m}{r}}-\frac{1}{E}),\sqrt{E^2-e^{\frac{-2m}{r}}},0,0)
\end{equation}

The metric in the new coordinates reads

\begin{equation}\label{geodmetric}
\displaystyle\mathrm{d}s^2=-\mathrm{d}\tau^2+\frac{E^2-e^{\frac{-2m}{r}}}{E^2}\mathrm{d}l^2+R_A^2(r)\mathrm{d}\Omega^2.
\end{equation}
We can see that $g_{ll}$ is finite and non-zero for any value of $r$, for $r=0$ the radial part of the metric even reduces to the Minkowski line element.

\begin{figure}[ht]\centering
	\includegraphics[scale=0.45]{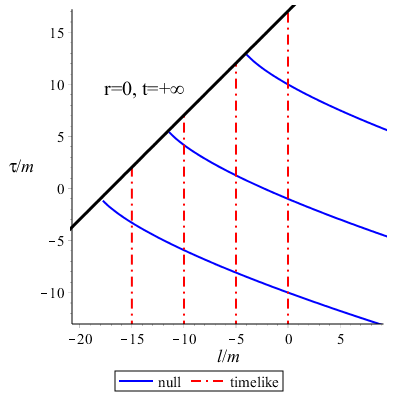}
\caption{Null and timelike geodesic observers arriving at $r=0$. The energy of timelike observers satisfies $E>1$, see \eqref{transformationform}.}
\label{ingoing}
\end{figure}

The transformation relations $x^\mu=(t,r)\rightarrow (\tau,l)$ can be found by solving

\begin{equation}\label{transformation}
\frac{\partial x^\mu}{\partial\tau}=U^\mu, \hspace{15pt} \frac{\partial x^\mu}{\partial l}=L^\mu.
\end{equation}
We now have to choose our geodesic observers. For $E>1$ the geodesics have no turning point and can reach $r=\infty $. We can further choose between ingoing observers coming from infinity to $r=0$ or outgoing ones going in the opposite direction. We then get the transformation in the form
\begin{subequations}\label{transformationform}
\begin{gather}
{t=\frac{\tau}{E}+f(\tau_\pm)},\hspace{15pt} r=R(\tau_\pm)\\
{\frac{\mathrm{d}f}{\mathrm{d}\tau_\pm}=E e^{\frac{2m}{R(\tau_\pm)}} -\frac{1}{E}}, \hspace{15pt} \frac{\mathrm{d}R}{\mathrm{d}\tau_\pm}=\pm\sqrt{E^2-e^{\frac{-2m}{R(\tau_\pm)}}},\hspace{10pt} \tau_\pm=\tau\pm l
\end{gather}
\end{subequations}
where the $+$ sign corresponds to the outgoing observers while $-$ to the ingoing ones. 
If we choose for example the ingoing observers we can depict the neighbourhood of $r=0$ better in these new coordinates than in $t,r$ as is illustrated  in figure \ref{ingoing}.

The null and timelike geodesics terminate at $\tau-l=R^{-1}(0)$ clearly displaying the geodesic incompleteness. At this point one would start to think about extending the metric beyond $r=0$ and it is really possible to smoothly extend it to Minkowski space if we could somehow discard the problematic angular part. \par
Even though the coordinates $(\tau,l)$ provide us with a new insight they fail to properly cover some parts of the spacetime. If we choose the ingoing observers in \eqref{transformationform} the set $(t=-\infty , r=0)$ is not well described in the new coordinates while for the outgoing observers the same can be said for $(t=+\infty , r=0)$, in both cases $(\vert t\vert < \infty , r=0)$ is  missing as it is inaccessible to timelike observers. 

There is yet another choice of geodesic observers which leads to new coordinates.  The geodesics with $E<1$ are travelling from $r=0$ to their turning point $r_{max}=-\frac{m}{\ln(E)}$ after which they arrive back to the origin. This coordinate chart then contains $(t=\pm\infty , r=0)$ but the part of the spacetime with $r>r_{max}$ is not covered at all.\par
Let us now briefly cover the non-radial motion, in particular the circular orbits. From the normalization of four-velocity $g^{\mu\nu}u_\mu u_\nu=\varepsilon$ (where $\varepsilon=-1$ or $\varepsilon=0$) we get the equation defining effective potential 

\begin{equation}\label{ur}
u_r^2 e^{-\frac {4m}{r}}=E^2-V_{eff}(r)
\end{equation}
where the effective potential itself is 

\begin{equation}\label{Veff}
V_{eff}(r)= \frac{{L}^{2}}{r^2} e^{-\frac {4m}{r}} -\varepsilon\,
{{\rm e}^{\,{-\frac {2m}{r}}}} .
\end{equation}

\begin{figure}[ht]\centering
\begin{subfigure}[b]{0.4\linewidth}
	\includegraphics[width=\linewidth]{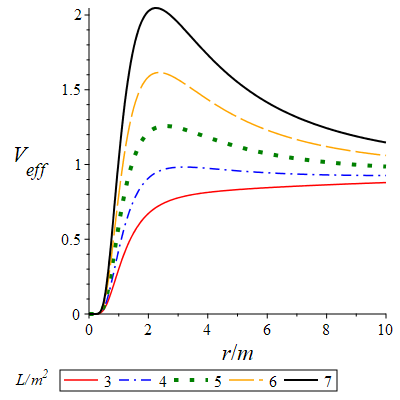}
	\caption{ }
	\end{subfigure}
	\begin{subfigure}[b]{0.4\linewidth}
	\includegraphics[width=\linewidth]{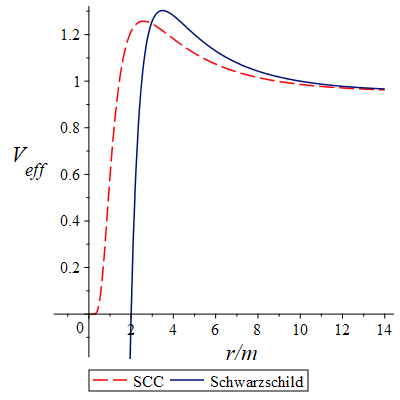}
	\caption{ }
	\end{subfigure}
\caption{(a) Plots of the effective potential $V_{eff}$ (in the case of massive particles) for several different values of the angular momentum $L$. (b) Comparison with the Schwarzschild  potential for a fixed value of $L= 5m^2$. }
\label{Vefffigure}
\end{figure}

For massive particles ( $\varepsilon=-1$) the potential looks similar to that of the Schwarzschild black hole. It has a stable and an unstable circular orbit for a sufficiently large angular momentum $L$. This can be seen in the figure \ref{Vefffigure} where  we can notice that the potential is everywhere finite and tends to zero for $r\rightarrow 0$.

In  order to find circular orbits we need to solve the equation $V_{eff}'(r)=0$. Here we will only mention the significant ones, the innermost stable circular orbit (ISCO) for the massive particles and the photon  unstable circular orbit in the massless case. While the photon sphere can be easily located the ISCO had to be found numerically

\begin{equation}\label{circ}
r_{photon}=2m, \hspace{35pt}  r_{ISCO} \approx 5.263m.
\end{equation}

{We can compare that to the Schwarzschild case where  $r_{photon}^{Sc}=3m$ and $r_{ISCO}^{Sc}=6m$.  At this point it is worth pointing out that the actual counterpart of  the Schwarzschild coordinate  $r$ is the areal radius defined in equation \eqref{radius} since it has the same geometrical meaning.  In this coordinate the values  are actually larger in our  spacetime than in Schwarzschild }

\begin{equation}\label{circ2}
R_{photon} \approx 3.297m, \hspace{35pt}  R_{ISCO} \approx 6.3643m.
\end{equation}

{While studying geodesics we can also compute the deflection angle for null geodesics to see how our spacetime behaves as a gravitational  lens. 
Similarly to the Schwarzschild case described in \cite{Iyer_2007}   we can calculate the angle by formula}

\begin{equation}\label{lensingSCC}
\delta\psi=2\displaystyle\int_{r_0(m)}^{\infty} {{\frac {b}{r\sqrt {{{\rm e}^{\,{\frac {4m}{r}}}}{r}^{2}-{b
}^{2}}}}} \mathrm{d}r -\pi
\end{equation}

{where $b=\frac{L}{E}$ is the impact parameter and $r_0(m)$ is the turning point of our null geodesic. We can expand this integral  to obtain a result for small masses (or large impact parameters)}

\begin{equation}\label{lensingSCCexpansion}
\delta\psi=\,{\frac {4m}{b}}+\,{\frac {4\pi \,{m}^{2}}{{b}^{2}}}+\mathcal{O}(m^3).
\end{equation}

{The same expansion done for Schwarzschild yields}

\begin{equation}\label{lensingSChwexpansion}
\delta\psi_{Schw}=\,{\frac {4m}{b}}+\,{\frac {15\pi \,{m}^{2}}{4{b}^{2}}}+\mathcal{O}(m^3).
\end{equation}
{We can see that while the first term is the same (and corresponds to the famous formula derived already by Einstein) the second term is slightly larger in our spacetime indicating modified lensing signature.}

\section{Conformal structure}
We shall now take a look at the global structure of the spacetime. In an approach similar to that used in the Schwarzschild spacetime we first express the metric in null coordinates using the tortoise coordinate $r^\ast(r)$

\begin{equation}\label{tortoise}
t^{\pm}=t\pm r^\ast(r),\hspace{15pt} r^\ast(r)= \displaystyle\int\limits_{m}^{r} e^{\frac{2m}{r^{'}}}\mathrm{d}r^{'}.
\end{equation}

we chose to have $r^\ast(m)=0$ as this is the location of the soon to be investigated wormhole throat (see the following section). The metric is then

 \begin{equation}\label{nullcoord}
\displaystyle\mathrm{d}s^2=-e^{-\frac{2m}{r}}\mathrm{d}t^{+} \mathrm{d}t^{-}+R_A^2(r)\mathrm{d}\Omega^2.
\end{equation}
At this point one would probably ask whether it is possible to have a finite and nonzero metric coefficient $g_{t^{+}t^{-}}$ at $r=0$. As the coefficient depends only on $r$ we have to use a transformation relations such that the multiplication factor produced by the transformation  is independent of $t$   while the metric remains in  null coordinates. This can be done (up to some multiplicative constants) using exponentials

 \begin{equation}\label{Kruskal}
u=\pm e^{\pm t^{+}},\hspace{15pt}  v=\mp e^{\mp t^{-}}.
\end{equation}
These are the familiar  null Kruskal coordinates. The metric then becomes

\begin{equation}\label{Kruskalmetric}
\displaystyle\mathrm{d}s^2=-e^{\mp 2r^\ast-\frac{2m}{r}}\mathrm{d}u \mathrm{d}v+R_A^2(r)\mathrm{d}\Omega^2.
\end{equation}
It is clear that we cannot make the radial part of the metric regular as $g_{uv}$ is either zero (the $+$ sign in \eqref{Kruskalmetric}) or diverges as $r\rightarrow 0^+$  ( the $-$ sign). \par
Despite this setback the metric in null coordinates is still useful for plotting the conformal diagram. It is easier to perform the compactification  using the form  \eqref{nullcoord}. Using the Kruskal coordinates \eqref{Kruskalmetric} directly would lead to the same picture with only the intervals of our new coordinates  $(T,R)$ being modified. For the compactification we use the same functions as in Minkowski spacetime {(here $[\cdot ,\cdot]$ denotes a closed interval)}

\begin{equation}\label{komapact}
t^{+}= \tan\left(\frac{1}{2}(T+R)\right), \hspace{15pt} t^{-}= \tan\left(\frac{1}{2}(T-R)\right), \hspace{15pt} T, R\in [-\pi,\pi].
\end{equation}
The metric in the new coordinates reads

\begin{equation}\label{metricTR}
\displaystyle\mathrm{d}s^2=\frac{e^{-\frac{2m}{r}}}{4 \cos^2\left(\frac{1}{2}(T+R))\cos^2(\frac{1}{2}(T-R)\right)}(-\mathrm{d}T^2 +\mathrm{d}R^2)+R_A^2(r)\mathrm{d}\Omega^2.
\end{equation}

The conformally conjugated metric can then be obtained as $ \widetilde{g}_{\mu\nu}=\Omega^2 g_{\mu\nu}$ where the conformal factor $\Omega$ is given by

\begin{equation}\label{faktor}
\Omega^2=4e^{\frac{2m}{r}} \cos^2\left(\frac{1}{2}(T+R)\right)\cos^2\left(\frac{1}{2}(T-R)\right).
\end{equation}

\begin{figure}[ht]\centering
	\includegraphics[width=55mm, height=50mm]{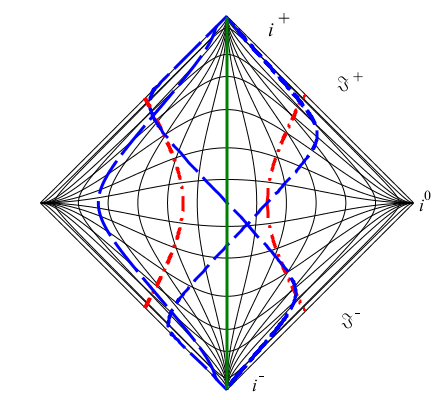}
\caption{The Penrose conformal diagram with timelike curves. The potential wormhole throat or the radius of minimal area ($r=m$) is marked in green (solid line), geodesics in blue (dashed line) and constantly accelerated observers in red (dash-dotted line).   }
\label{Penrose1}
\end{figure}

\begin{figure}[ht]\centering
	\includegraphics[width=55mm, height=50mm]{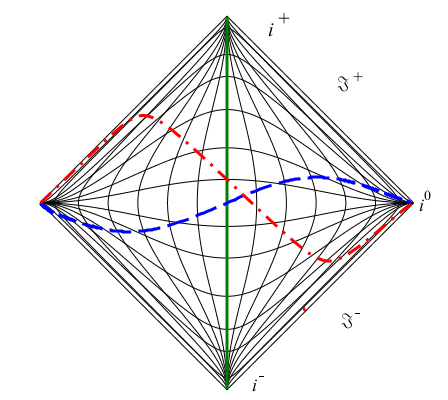}
\caption{The Penrose conformal diagram with spacelike curves. The wormhole throat ($r=m$) is marked in green (solid line),  a spacelike geodesic in blue (dashed line) and a $\tau= const.$ curve  (see \eqref{transformationform})  in red (dash-dotted line).  }
\label{Penrose2}
\end{figure}

It can be shown that the conformally related metric $ \widetilde{g}_{\mu\nu}$ has all coefficients finite. The boundary of the physical spacetime is of course given by $\Omega=0$ which is also the boundary of the corresponding conformal diagram as depicted in figures \ref{Penrose1} and \ref{Penrose2}.

{When looking at the diagrams in figures \ref{Penrose1} and \ref{Penrose2} we notice that they look like two copies of the Minkowski spacetime  compactified using spherical coordinates $(t,r)$ and glued together along $r=0$. On the  right side  we have the usual structure of the asymptotically flat spacetime with infinities $i^0$, $i^\pm$ and $ \mathcal{I^\pm}$. Since the left side boundary looks indistinguishable one would think that it could also have its own space and null infinities.  As was already stated the spacetime can be interpreted as a wormhole and indeed the conformal diagram looks like that of a  static spherically symmetric wormhole connecting two asymptotically flat regions (see for example \cite{Simpson_2019} or \cite{HAYWARD_1999}).}

The figure  \ref{Penrose1} features all three types of geodesics (ingoing, outgoing and those with $E<1$) while it also includes two constantly accelerated (Rindler-like) observers ($a_{\mu} a^{\mu}=const.$). The way the worldlines approach infinity is the same at both sides of the diagram (apart from the $E<1$ geodesics) but what is really different from an ordinary wormhole is the asymmetry with respect to its throat which is marked in green. The motion on the left side of the diagram ($r<m$) takes only a finite amount of proper time. This was already mentioned in the previous section but what is really unintuitive here is the fact that the future time infinity $i^+$ can be reached in a finite proper time provided that we travel there along a geodesic that approaches it from the left (the same obviously applies for  $i^-$). 

This means that the region inside the wormhole throat works as a one-directional time machine enabling travel to infinitely distant future (as perceived by observers outside of the wormhole throat) in finite time by entering the wormhole for certain time. 

This strange finiteness is present in the case of spacelike curves as well. Some of the spacelike curves can be seen in figure \ref{Penrose2}. While the spacelike geodesics $t=const.$ (in black) reach the "left spatial infinity" (the point $(\vert t \vert <\infty ,r=0)$) after an infinite distance ( see  \eqref{properdistance}) the radial geodesics not confined in a $t=const.$ plane  need to cover only a finite distance to reach the point. The same can be also said for the  coordinate line $\tau=const.$ in the comoving coordinates \eqref{transformationform}. This is again valid only for the left side and the distance to right $i^0$ is obviously infinite no matter which curve we choose.\par 
As was already stated above the conformal structure looks identical to that of the Minkowski spacetime and so it is not surprising that the spacetime is asymptotically simple which means that $\Omega$ satisfies \cite{bojowald_2010} 

\begin{equation}\label{assimple}
\mathrm{d}\Omega( \mathcal{I^\pm})\neq0, \hspace{35pt}   \mathrm{d}\Omega(i^0)=0.
\end{equation}
The same is surprisingly  true also for the infinities on the left. On the other hand the spacetime is not asymptotically empty as it is filled with the phantom scalar field (the condition $T_{\mu \nu}=\mathcal{O}(\Omega^3)$ given in \cite{bojowald_2010} is not satisfied). It is thus not correct to call the spacetime asymptotically flat even though it possesses the same asymptotic structure.\par
\begin{figure}[ht]\centering
\begin{tikzpicture}[scale=0.6]
\node (I)    at ( 4,0)   {};
\path % Four conners of the right diamond (no labels this time)
   (I) +(90:4)  coordinate[label=90:$i^+$] (Itop)
       +(-90:4) coordinate[label=-90:$i^-$] (Ibot)
       +(180:4) coordinate (Ileft)
       +(0:4)   coordinate[label=0:$i^0$] (Iright)
       ;
% No text this time in the next diagram
\draw  (Itop) --
       node[midway, above right]    {$\mathcal{I}^+$}  
 (Iright) -- 
 node[midway, below right]    {$\mathcal{I}^-$}  
 (Ibot);
\draw[decorate,decoration=zigzag] (Itop) -- (Ileft)
      node[midway, above, outer sep=5mm] {$r=0$};

\draw[decorate,decoration=zigzag] (Ileft) -- (Ibot)
      node[midway, below, outer sep=5mm] {$r=0$};       
\draw[dashed] (Itop) -- (Ibot)
 node[midway,right] {$r=m$};    
\end{tikzpicture}
\caption{Schematically  drawn  Penrose conformal diagram  including the singularity, the throat and the asymptotic infinities } \label{Penrosesingular}
\end{figure}
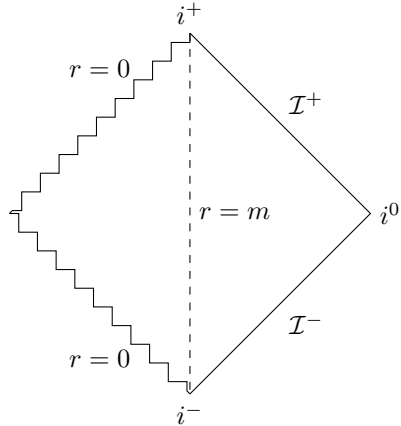

Apart from  the finite lengths of the worldlines in the $r<m$ part the spacetime looks like  a  generic static wormhole \cite{HAYWARD_1999}. The biggest difference  however is the presence of a curvature  singularity in $r=0$ which will be discussed further in the section \ref{section7} of this paper. The conformal structure of the spacetime is then depicted in the figure \ref{Penrosesingular}.

\section{The wormhole}

The wormhole structure of the spacetime becomes evident when studying null congruences. The two independent future-pointing radial null vectors (an outgoing $k$ and an ingoing $l$ )   can be written as
\begin{equation}\label{kongruence}
k^\mu=k^t(x^\mu)(1, e^{-\frac{2m}{r}},0,0) ,\hspace{15pt}  l^\mu=\frac{1}{k^t(x^\mu)} (e^{\frac{2m}{r}} ,-1,0,0)
\end{equation}
where $k^t(x)>0$ is a  function of all coordinates while the vectors satisfy the normalization condition $k^\mu l_\mu =-2$. Using the metric on the sphere $\sigma_{\mu\nu}$ we can compute the expansion using the formula

\begin{equation}\label{expanzedefinice}
\theta^{(k)}=\sigma^{\mu\nu} \triangledown_{\mu} k_\nu.
\end{equation} 

The resulting expansions then read

\begin{equation}\label{expanze}
\theta^{(k)}=\frac {2 \left( r-m \right) {\it k^t(x^\mu)  e^{-\,{
\frac {2m}{r}}}}}{{r}^{2}}
 ,\hspace{15pt}  \theta^{(l)}=-{\frac {2(r-m)}{{r}^{2}{\it k^t(x^\mu)}}}.
\end{equation} 

For $r>m$ the outgoing and the ingoing expansion have expected  signs $\theta^{(k)}>0$ and $\theta^{(l)}<0$. This changes at the minimum of the areal radius $R_A$ located at $r=m$ where the expansions vanish and then flip signs for $r<m$. This is consistent with the notion of a  wormhole  with a throat at $r=m$. \par 
In order to obtain a better insight it is useful to compare our spacetime to a more simple one such as the Bronnikov--Ellis (BE) wormhole \cite{Elliswormhole} which is the simplest static spherically symmetric wormhole given by metric

\begin{equation}\label{BEwormhole}
\displaystyle\mathrm{d}s^2_{BE}=-\mathrm{d}t^2 + \mathrm{d}x^2+({x}^{2}+{m}^{2})\mathrm{d}\Omega^2.
\end{equation} 

The convenience  of the  coordinates used in the BE wormhole stems from the form of the areal radius ($g_{\theta\theta}$) which is a symmetric function  with respect to the throat at $x=0$.
In our case we can also find analogous coordinate $x \in (-\infty,\infty )$  which can be defined as

\begin{equation}\label{newcoordinate}
x(r)={\mathrm{signum}} \left(r-m \right) \sqrt {{{\rm e}^{\,{\frac {2m}{r}}
}}{r}^{2}-{m}^{2}{{\rm e}^{2}}}.
\end{equation}

The  inverse transformation is then

\begin{equation}\label{inverse}
r(x\neq 0)=-{ \frac{m\,{ \Theta}\left( x \right) }{   \ { W} \left(-\sqrt {{
\frac {{m}^{2}}{{x}^{2}+{m}^{2}{{\rm e}^{2}}}}}\right) } }-
{\frac { m\,{\Theta} \left( -x \right) }{{W} \left(-1,-\sqrt {
  \frac{m^2}{\left( {x}^{2}+m^2{{\rm e}^{2} } \right)}}\right)}};\hspace{15pt} r(0)=m
\end{equation} 
where $\Theta (x)$ is the step function while  $W(x)$ and $W(-1,x)$ are the branches of Lambert $W$-function. The metric then has the desired form

\begin{equation}\label{whmetric}
\displaystyle\mathrm{d}s^2=g_{tt}(x)\mathrm{d}t^2 + g_{xx}(x)\mathrm{d}x^2+({x}^{2}+{m}^{2}e^2)\mathrm{d}\Omega^2.
\end{equation} 

\begin{figure}[ht]\centering
	\begin{subfigure}[b]{0.4\linewidth}
	\includegraphics[width=\linewidth]{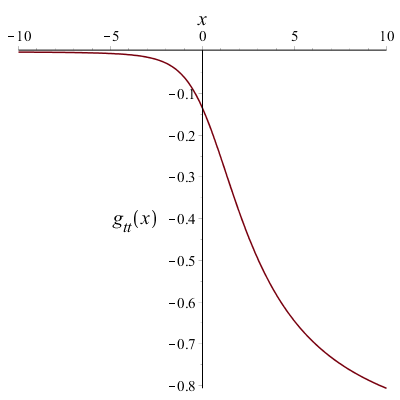}
	
	\end{subfigure}
	\begin{subfigure}[b]{0.4\linewidth}
		\includegraphics[width=\linewidth]{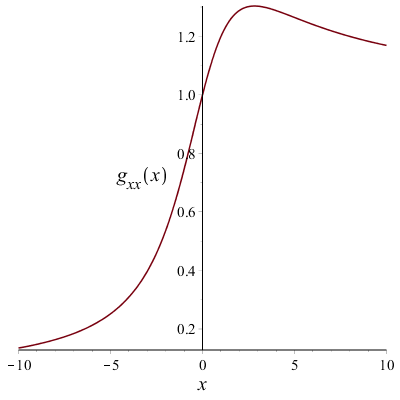}
		
	\end{subfigure}
\caption{The  functions  $g_{tt}(x)$ (left) and $g_{xx}(x)$ (right) of the metric \eqref{whmetric} for $m=1$.}
\label{whcoord}
\end{figure}

The functions $g_{tt}(x)$ and $g_{xx}(x)$ are rather lengthy expressions and so it serves no purpose to actually write them down here when it is easy to plot them (see figure \ref{whcoord}).

It is clear from the plots that the metric functions are not symmetric with respect to $x=0$ with the metric \eqref{whmetric} approaching Minkowski for $x\rightarrow \infty$ while for $x\rightarrow -\infty$ both $g_{tt}(x)$ and $g_{xx}(x)$ vanish.\par 

When studying wormholes it is common to perform embedding of the plane $t=const., \theta=\frac{\pi}{2}$ to illustrate how two asymptotically flat universes are connected by its funnel-like structure. In our case the cylindrical coordinate $\rho$ in the euclidean space is connected to the coordinate $r$ as
\begin{equation}\label{euclidr}
\rho= r{{\rm e}^{\,{\frac {m}{r}}}}
\end{equation} 
which is in fact the already mentioned areal radius. To find the embedding diagram $z(\rho)$ (z also being an euclidean coordinate) we can easily derive the well known formula

\begin{equation}\label{embed1}
\frac{\mathrm{d}z}{\mathrm{d}\rho}=\pm \sqrt{g_{\rho\rho}(\rho)-1}.
\end{equation} 

\begin{figure}[ht]\centering
	\begin{subfigure}[b]{0.4\linewidth}
	\includegraphics[width=\linewidth]{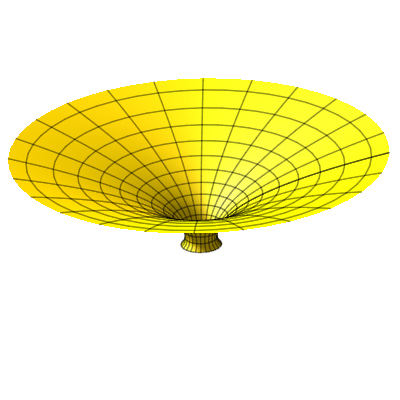}
	
	\end{subfigure}
	\begin{subfigure}[b]{0.4\linewidth}
		\includegraphics[width=\linewidth]{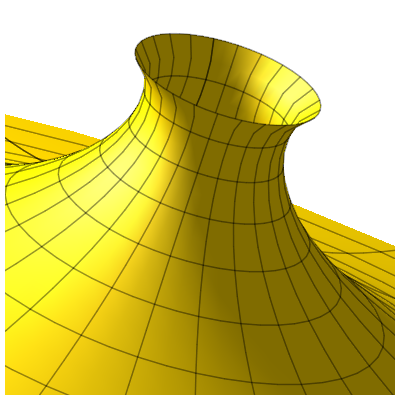}
		
	\end{subfigure}
\caption{Embedding of the equatorial plane into the Euclidean space}
\label{embed}
\end{figure}

To get the function $g_{\rho\rho}$ one must invert the relation \eqref{euclidr}. However, this is only possible for each interval $r>m$ or $r<m$ separately. So instead of applying \eqref{embed1} globally we use different $g_{\rho\rho}$ for each interval
\begin{subequations}\label{embed2}
\begin{gather}
\label{embed2a}
\frac{\mathrm{d}z}{\mathrm{d}\rho}=\sqrt{g^{(1)}_{\rho\rho}(\rho)-1} \hspace{35pt} r>m\\
 \label{embed2b}
\frac{\mathrm{d}z}{\mathrm{d}\rho}=-\sqrt{g^{(2)}_{\rho\rho}(\rho)-1} \hspace{35pt} r<m.
 \end{gather}
\end{subequations} 
The functions $g^{(i)}_{\rho\rho}$ on the respective intervals can again be expressed using the Lambert functions. This allows us to compute $z(\rho)$ which in turn can be used to plot the embedding diagram in the figure \ref{embed}.

When looking at the diagram we can once again see the asymmetry with respect to the throat. The embedded surface gradually approaches the plane $z=0$ in the limit $r\rightarrow \infty$ which is an expected behaviour. On the other side of the throat however the diagram ends abruptly before reaching $r=0$ (at the point $\rho_0$ for which $g^{(2)}_{\rho\rho}(\rho_0)=1$). This can be interpreted by saying that after passing through the throat the space expands so quickly  that it no longer fits into the euclidean space. This is closely connected to the problem of reaching $i^+$ via the $r<m$ region in a finite time which was discussed above.\par
So far we have been describing our spacetime as if it were a wormhole without proving that it satisfies  any formal definition of  wormhole. Let us now briefly examine two such definitions both of which define wormholes as spacetimes with a throat but without an event horizon in the vicinity of the throat. The definitions differ in their conditions for a throat. The first definition, due to Morris and Thorne \cite{Thorne}, employs the equation of the embedded surface $z=z(\rho)$ and states that at the throat $\rho =\rho_0$ the following condition are to be satisfied

\begin{subequations}\label{thornedef}
\begin{gather}
\label{eq:thornedefa}
\displaystyle{\lim_{\rho \to\rho_0}} \frac{\mathrm{d}z}{\mathrm{d}\rho}=\infty \\
 \label{eq:thornedefb}
 \displaystyle{\lim_{\rho \to\rho_0}} \frac{\mathrm{d^2}\rho}{\mathrm{d}z^2}>0 .
 \end{gather}
\end{subequations} 

Both of these conditions of course apply only for  the positive branch of $z(\rho)$ in \eqref{embed1} as both of them are identical in case of the Morris-Thorne wormhole (determined by the shape function $b(\rho)$ in \cite{Thorne}). 
Our spacetime  is not symmetric with respect to the throat at $\rho_0=m e$ so we have to verify the conditions for each branch separately. The first condition clearly holds true as can be seen from the embedding diagrams.
 The second condition (eq. \eqref{eq:thornedefb}) is the so called flare-out condition which is again satisfied with an identical result for both branches
  
 $$\displaystyle{\lim_{\rho \to\rho_0}} \frac{\mathrm{d^2}\rho}{\mathrm{d}z^2}= 2 m e.$$
  
An alternative definition of wormhole was introduced by Visser \cite{Visser}. It defines a wormhole  throat as a spacelike surface with two null congruences normal to it. In our case it is the surface $t=const., r=m$ with congruences $k$ and $l$ as defined in \eqref{kongruence}. The expansion scalar should then satisfy

 \begin{subequations}\label{Visser}
\begin{gather}
\label{vissera}
\theta^{(k)}\vert_{r_0}=0\\
 \label{visserb}
\dot{\theta}^{(k)}\vert_{r_0}= \pounds_{k} \theta^{(k)}\vert_{r_0}\geq 0 .
 \end{gather}
\end{subequations} 
The second equation is again called flare-out condition. The same conditions are also supposed to hold true for the congruence $l$. Using the expressions \eqref{expanze} one can easily check that both conditions are satisfied. It is not even necessary to compute the derivatives one can simply use the Raychaudhuri equation for null geodesics

\begin{equation}\label{Raychaudhuri}
\dot{\theta}=-\frac{1}{2}\theta^2-\sigma_{\mu\nu}\sigma^{\mu\nu} +\omega_{\mu\nu}\omega^{\mu\nu}- T_{\mu \nu} k^\mu k^\nu.
\end{equation} 
Since the expansion at the throat vanishes and the other optical scalars are identically zero we are left with

\begin{equation}\label{Raychaudhuri2}
\dot{\theta}^{(k)}\vert_{r_0}=-( T_{\mu \nu} k^\mu k^\nu) \vert_{r_0}>0
\end{equation} 
which is just a consequence of the violation of the null energy condition which is a generic feature of static wormholes \cite{Gibbons:2017jzk}.
As the absence of an event horizon in our spacetime is evident we can conclude that it is indeed a wormhole.

\section{Mass-energy of the spacetime}

We would now like to examine the mass/energy content of the spacetime.  Since the spacetime has the same structure as an asymptotically flat one the notion of total mass is well defined. Taking advantage of the existence of Killing vector $\xi^{(t)}_{\nu}$ we can easily compute the Komar mass \cite{Komar} as a surface integral over a sphere located at the infinity

\begin{equation}\label{Komar}
M_K=-\frac{1}{8\pi}\displaystyle\int_{S\rightarrow \infty} \triangledown_{\mu} \xi^{(t)}_{\nu} \mathrm{d}S^{\mu\nu}=\frac{1}{4\pi}\displaystyle\int_S m \mathrm{d}\Omega=m 
\end{equation} 
where { $\mathrm{d}S_{\mu\nu}=\sqrt{-g}\, \mathrm{d}\theta \,\mathrm{d}\phi\hspace{5pt} \mathrm{d}t\wedge\mathrm{d}r$}. The result is $m$ and gives an expected interpretation to the metric parameter. What may be confusing is that the total mass is positive while the spacetime is filled with a scalar field with negative energy density. However, one must realize that the result is in perfect accordance with the behavior of geodesic observers who are attracted to  $r=0$ which would not be the case if $M_K<0$. This leads us to conclude that the negative energy of the scalar field is somehow compensated by positive energy of the gravitational field. In order to study this further we can try to use several concepts of quasi-local energy to find out how much energy/mass is located inside a sphere of a given radius $r=const.$ (finding the function $E(r)$). \par
We can obviously start from the Komar energy but it turns out that when evaluating the expression  \eqref{Komar} for an arbitrary sphere the result is the same ($m$).

Another concept of energy that can be applied is the Brown--York (BY) energy which can be obtained from the canonical  (ADM) formalism (derived for example in \cite{bojowald_2010}). It is derived based on the on-shell boundary contribution to the total hamiltonian and provides evolution hamiltonian for canonical boundary data (the bulk constraints vanish for a solution of Einstein equations). We can again compute it as an integral over a sphere

 \begin{equation}\label{BY}
E_{BY}=-\frac{1}{8\pi} \displaystyle\int_S (k-k_0) \sqrt{\mathrm{det}\sigma}\,\mathrm{d}^2x
 \end{equation}
where $\sigma$ is the restriction of the metric to the sphere $(\sigma=g\vert_{S})$ and $k$ is the trace of extrinsic curvature of the sphere (as a slice in a spatial hypersurface) while $k_0$ refers to the extrinsic curvature of the same sphere as  embedded in the flat space which ensures that $E_{BY}=0$ for the Minkowski spacetime. If we compute the BY energy we indeed obtain a non-trivial function
$E_{BY}(r)$ but with a wrong asymptotics  $E_{BY}(\infty)=M_{BY}=2m$. This result is clearly not consistent with the Komar mass \eqref{Komar}. To understand this discrepancy let us first move to next concept of energy. The Brown--York mass is often confused with the ADM mass which is given by

 \begin{equation}\label{MADM}
M_{ADM}=\frac{1}{16\pi}  \displaystyle\int_{S\rightarrow \infty}(D^i \gamma_{ij}- D_j \gamma)r^j \sqrt{\mathrm{det}\sigma}  \,   \mathrm{d}^2x
 \end{equation}
where $\gamma$ is the difference between the spatial part of the metric ($h$) of the spacetime in question and the spatial part of the Minkowski metric ($h^{(0)}$) given in the same type of coordinates ($\gamma_{ij}=h_{ij}-h^{(0)}_{ij}$). The derivative $D$ is the flat-space covariant derivative and $r^j$ is the normal to the sphere. The ADM mass of course gives the correct result $M_{ADM}=m$. Brown--York mass coincides with the ADM mass only if the metric restricted to our sphere is the same as in the reference flat space i. e.  $\gamma_{ij}\rvert_S=0$ (as was shown in \cite{Hawking_1996} ). We are thus forced to compute the BY mass in coordinates in which $r$ is replaced by the areal radius according to the transformation \eqref{euclidr}. If we use this transformation we then obtain the correct result for mass but as in the case of the Komar mass we get $E_{BY}(\rho)=m$ which does not give us any information about how the energy is distributed in the spacetime.\par
The final concept of energy we use is the Misner--Sharp energy \cite{MS} which  can be defined  as 
  \begin{equation}\label{EMS}
E_{MS}(r)=\frac{1}{8} R_A^3 R_{\mu\nu\rho\sigma} \epsilon^{\mu\nu}\epsilon^{\rho\sigma}, \hspace{35pt} \epsilon_{\mu\nu}=\epsilon_{\mu\nu\rho\sigma}n^\rho r^\sigma
 \end{equation} 
where $n^\rho$ is the normal to $t=const.$ hypersurface while $r^\sigma$ is a spatial normal to the sphere. This energy can only be used in the spherically symmetric case and it again tells us how much energy is located inside a sphere given by the coordinate $r$. Application to the SCC spacetime gives the following result
   \begin{equation}\label{EMSCC}
E_{MS}(r)=-\,\frac {m \left( m-2\,r \right) e^{{\frac {m}{r}}}}{2r}.
 \end{equation} 
 This function has the correct asymptotic behaviour $E_{MS}(\infty)=m$ and in contrast to the previous energies it is a non-constant function. By taking the limit $r\rightarrow 0^+$ we can see that the energy located at $r=0$ is $-\infty$ which again supports the presence of the singularity there and it is in fact not that surprising since the scalar field is divergent there. 
 
 We can also compute the total energy of the scalar field itself from its stress-energy tensor which has a familiar form
   \begin{equation}\label{Tscalar}
T_{\mu\nu}=-\left(\nabla_\mu \Phi \nabla_\nu \Phi-\frac{1}{2}  g_{\mu\nu} (\nabla\Phi)^2 \right).
 \end{equation}  
 {Let us note that this stress-energy tensor clearly violates null energy condition (and therefore all the stronger ones) everywhere in our spacetime.}
  To restore the gravitational constant $\kappa=8\pi$ which was absorbed into the scalar field in the action \eqref{action} we compute the energy for the scalar field $\Phi\rightarrow\sqrt{\frac{2}{\kappa}} \Phi$
\begin{equation}\label{Escalar}
E_{scalar}=\displaystyle\int_{t=const.} n_\mu T^{\mu\nu}n_{\nu}\sqrt{h} \mathrm{d}^3x=\displaystyle\int_0^{\infty} -\frac{m^2 e^{{\frac {m}{r}}}}{r^2} \mathrm{d}r=-\infty.
 \end{equation} 
This gives us again an infinite result due to the singular behaviour in $r=0$.  We would now like to avoid the infinities in the energies and instead switch to the energy densities. For the scalar field (as for any matter field) we already have an explicit formula while the total energy density is in principle obtained from $E_{MS}(r)$.  There is however a slight complication as we have $E_{MS}(0)=-\infty$ in contrast to the usual $E_{MS}(0)=0$. This strange behaviour leads us to define the density using an integration from the actual zero point of $E_{MS}(r)$ at $r=\frac{m}{2}$
 
\begin{equation}\label{MSdensity}
E_{MS}= \displaystyle\int^{\infty}_{\frac{m}{2}} \varrho_{MS}(r) 4\pi\sqrt{g_{rr}g^2_{\theta \theta}}  \mathrm{d}r.
 \end{equation} 
 
The densities can then be computed using the formulas

   \begin{equation}\label{Edensitydef}
\varrho_{MS}(r)=\frac{1}{4\pi\sqrt{g_{rr}g^2_{\theta \theta}}}\frac{\mathrm{d} E_{MS}(r)}{\mathrm{d}r},  \hspace{20pt}  \varrho_{scalar}(r)= n_\mu T^{\mu\nu}n_{\nu}
 \end{equation} 
 and after the evaluation we get
 
    \begin{equation}\label{Edensity}
\varrho_{MS}(r)=-\frac{1}{8\pi}\,{\frac {{m}^{2} \left( r-m \right) }{{r}^{5}}{{\rm e}^{-\,{
\frac {2m}{r}}}}}
,  \hspace{20pt}  \varrho_{scalar}(r)= -\frac{1}{8\pi}\,{\frac {{m}^{2}}{{r}^{4}}{{\rm e}^{-\,{\frac {2m}{r}}}}}.
 \end{equation} 

\begin{figure}[ht]\centering
	\begin{subfigure}[b]{0.4\linewidth}
	\includegraphics[width=\linewidth]{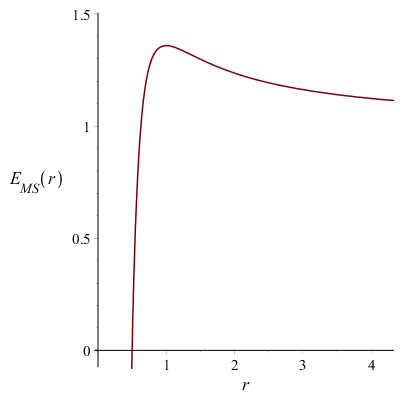}
	\caption{}
	\end{subfigure}
	\begin{subfigure}[b]{0.4\linewidth}
		\includegraphics[width=\linewidth]{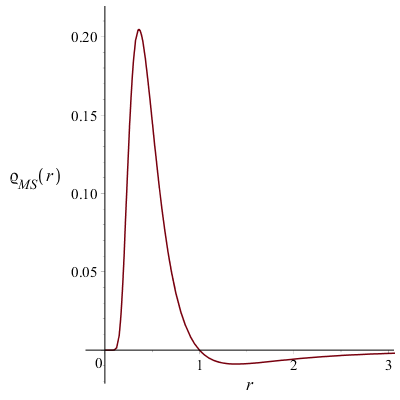}
	\caption{}	
	\end{subfigure}
\caption{The Misner--Sharp energy  $E_{MS}(r)$ and the corresponding energy density $\varrho_{MS}(r)$ for $m=1$. }
\label{energy}
\end{figure}

  When looking at the figure \ref{energy} depicting Misner--Sharp energy and energy density we can again see the asymmetry with respect to the throat. For $r<m$ the total energy density is positive while in the asymptotically  flat  (or rather simple) part the energy density is negative.\par

As was stated above the negative energy of the scalar field should be somehow compensated by the positive energy of the gravitational field. To get some further insight into it we would like to find the gravitational  energy density. We are fully aware of the ambiguity of the definition of this the concept. Often defined in terms of various pseudo-tensors there are disputes whether any meaningful definition of localized gravitational energy even exists. However, in our simple case of a spherically symmetric spacetime there is a way how to define a notion of gravitational energy density with reasonable properties. The naive approach would be to simply subtract the scalar field  energy density from the total (in our case Misner--Sharp) energy density
 \begin{equation}\label{gravdensity}
\varrho_{gravity}(r)=\varrho_{MS}(r)-\varrho_{scalar}(r)=\frac{1}{8\pi}\,{\frac {{m}^{3}}{ \,{r}^{5}}{{\rm e}^{-\,{\frac {2m}{r}}}}}.
 \end{equation} 

There is however a principal problem with this straightforward approach. Although the volume integral of a quantity over a given region of space gives us the total gravitational energy it does not mean that the quantity is the gravitational energy density. In the Newtonian theory the correct expression for energy density is $\frac{1}{2} E^2$ not $\frac{1}{2} \rho \Phi$. This was correctly pointed out by Katz, Lynden-Bell and Bičák (KLB) in \cite{LyndenBell2007EnergyAA} who managed to find a prescription for the energy density in the form of a quadratic expression as in the Newtonian case. They considered a spherically symmetric metric in the form
\begin{equation}\label{KLBmetric}
\displaystyle\mathrm{d}s^2=-\xi^2(\rho)\mathrm{d}t^2+\frac{1}{1-\frac{2m(\rho)}{\rho}}\mathrm{d}\rho^2+\rho^2 \mathrm{d}\Omega^2,
\end{equation}
where the function $m(\rho)$ gives us the total energy inside a sphere with the areal radius $\rho$. It can be shown that the function  $m(\rho)$ is in fact the Misner--Sharp energy of the spacetime. The metric \eqref{KLBmetric} uses the areal radius $\rho$ as a coordinate but in our case we have two different spheres with the same radius $\rho$. This was already discussed in the previous section concerning the wormhole embedding (see equation \eqref{embed2}).  We will now restrict ourselves to the region $r>m$ which has a regular behaviour at the infinity. Subsequently, we will briefly describe the situation for $r<m$ .\par 

KLB managed to find their expression for gravitational energy density in the desired form $\frac{1}{2} F^2$ where in analogy with Newtonian theory  they define gravitational intensity $F$ with a corresponding potential  $\Phi$
  
\begin{equation}\label{intensitypotential}
F=-\frac{1}{\rho}\bigg[1-\bigg(1-\frac{2m(\rho)}{\rho}\bigg)^{\frac{1}{2}}\bigg], \hspace{15pt} F=-(g_{\rho\rho})^{-\frac{1}{2}} \frac{\mathrm{d}\Phi}{\mathrm{d}\rho}
\end{equation} 
For convenience we added a minus sign to the original KLB definition of $F$. The resulting intensity and potential in the region  $r>m$ are then

\begin{equation}\label{intensitypotential1}
F^{(1)}=-{\frac {m}{{r}^{2}}{{\rm e}^{-{\frac {m}{r}}}}}, \hspace{15pt} \Phi^{(1)}=-\frac{m}{r}.
\end{equation}
This result can be well-interpreted in the Newtonian perspective. If we write down the Newtonian limit of the geodesic equation (for a radial geodesic) we find that the radial component of four-velocity satisfies
\begin{equation}\label{geodesicnewton}
\frac{\mathrm{d}u_r}{\mathrm{d}t}=-{\frac {m}{{r}^{2}}{{\rm e}^{-{\frac {m}{r}}}}}+\mathcal{O}(p_r^2)
\end{equation} 
This was expected as the potential $ \Phi^{(1)}$ is also the gravitational potential in the Newtonian limit. Naturally the gravitational energy density is then
 \begin{equation}\label{KLBdensity}
\varrho^{(1)}_{KLB}(r)=-\frac{1}{8\pi}\frac{\mathrm{d}\Phi^{(1)}}{\mathrm{d}\rho} g^{\rho\rho}\frac{\mathrm{d}\Phi^{(1)}}{\mathrm{d}\rho}=-\frac{1}{8\pi} \left(F^{(1)}\right)^2=-\frac{1}{8\pi}\,{\frac {{m}^{2}}{{r}^{4}}{{\rm e}^{-\,{\frac {2m}{r}}}}}.
\end{equation} 
Following the results  of KLB we can apply the spatial Laplace operator on $ \Phi^{(1)}$

  \begin{equation}\label{Laplace}
\triangle \Phi^{(1)} = \frac{1}{r^2 e^\frac{3m}{r}} \frac{\mathrm{d}}{\mathrm{d}r}\bigg ( r^2 e^\frac{m}{r}\frac{\mathrm{d}}{\mathrm{d}r}\Big(-\frac{m}{r}\Big)\bigg)=-{\frac {{m}^{2}}{{r}^{4}}{{\rm e}^{-\,{\frac {2m}{r}}}}}
 \end{equation} 
 
and see that the Poisson equation is satisfied 
 
 \begin{equation}\label{poisson}
\triangle \Phi^{(1)} = 4\pi ( \varrho_{scalar} + \varrho^{(1)}_{KLB}).
 \end{equation}  
Here we can see that the gravitational energy \eqref{KLBdensity} contributes to the right hand side of \eqref{poisson} the same as the energy of the scalar field \eqref{Edensity}. \par 
It is important to remark here that for a general spherically symmetric spacetime the potential  $\Phi$ does not reduce to the gravitational potential in the Newtonian limit. As  KLB found out  the potential has a geometric meaning,  $e^{2\Phi}$ is a conformal factor of a transformation between the spatial part of the metric and the flat space.\par 
 On the other hand, the Newtonian gravitational potential is given by the metric component $g_{tt}$ but none of the quantities derived by KLB depend on the function  $\xi(\rho)$. The fact that the potential  $ \Phi^{(1)}$ not only satisfies the Poisson equation but also plays the role of gravitational potential in the geodesic equation is given solely by the particular form of our metric (or more generally by the metric \eqref{Gibbons}).
 
We can now turn to the region   $r<m$.  Using the same definition \eqref{intensitypotential} as in $r>m$ we obtain the respective intensity along with the corresponding potential

 \begin{equation}\label{intensitypotential2}
F^{(2)}={\frac {m-2\,r}{{r}^{2}}{{\rm e}^{-{\frac {m}{r}}}}}, \hspace{15pt} \Phi^{(2)}=-\frac{m}{r}-2\ln(r) .
\end{equation}

As in the previous case this potential satisfies the Poisson equation \eqref{poisson} but other than that this result has some strange properties. The first thing one may notice is the fact that both the intensity and the potential do not vanish for $m\rightarrow 0$. This may be the consequence of the usage of the coordinate $\rho$ ( the areal radius \eqref{euclidr}) in the region $r<m$. The metric in these coordinates does not converge  to the Minkowki metric when $m\rightarrow 0$ because the region $r<m$ gradually shrinks and then disappears in the limit $m\rightarrow 0$. \par 
Another important aspect is the fact that $F^{(2)}$ flips sign at $r=\frac{m}{2}$. This is consistent with the behaviour of the Misner--Sharp energy \eqref{EMSCC} which also changes sign there. However, this is not reflected by the geodesic motion. The point  $r=\frac{m}{2}$ is not a local minimum of the effective potential \eqref{Veff} while the test particles are attracted to $r=0$ whatever their location. We can thus conclude that the behaviour of the intensity, potential, as well as the  Misner--Sharp energy in this region is very non-intuitive and cannot be interpreted from a Newtonian viewpoint.

Apart from the issues described in the previous paragraph the main problem of the KLB approach is the fact that the integral of the total energy density fails to give us the total mass $m$  
 
     \begin{equation}\label{perpartes}
\displaystyle\int^{\infty}_{\frac{m}{2}} (\varrho_{scalar}(r)+\varrho_{KLB}(r) )4\pi\sqrt{g_{rr}g^2_{\theta \theta}}  \mathrm{d}r \neq m.
 \end{equation} 
 
 As already mentioned above the negative energy of the scalar field should be compensated by a positive energy of the gravitational field. But the  density $\varrho_{KLB}(r)$ is defined with a negative sign (equation \eqref{KLBdensity}) so it is clear that we cannot get a positive total mass $m$. When KLB derived their expression for energy density they subtracted the matter energy from the total (Misner--Sharp) energy. After that they performed an integration by parts where the surface term vanished. This surface term is however non-zero in our spacetime which is the reason that the integral  \eqref{perpartes} is not equal to $m$.
 
 To summarize, the advantage of the KLB approach is its Newtonian interpretation which in our spacetime also includes the potential in the  geodesic equation. On the other hand the integral \eqref{perpartes} is not equal to the total mass/energy of the spacetime. The naive  gravitational energy density   \eqref{gravdensity} obtained as a simple difference has exactly opposite problems.

\section{The non-scalar singularity}\label{section7}

The original motivation when deriving the metric \eqref{scc} was to eliminate the direction-dependent curvature singularity of the original Curzon--Chazy spacetime.  It turns out that the Kretschman scalar \eqref{Kretschmann} is finite if the metric function $k$  vanishes so this seems to be the right path to a singularity-free spacetime. Not only the Kretschmann scalar but also other invariants constructed from the Riemann tensor are finite which holds also for the Newman-Penrose scalars $\Psi_a$ and  $\Phi_{ab}$. \par 
On the other hand as was shown in the section 3  the spacetime is geodesically incomplete, the null hypersurface  $r=0$ can be reached in a finite proper time. A question stands however. Can the spacetime be extended beyond  $r=0$? In the original  Curzon--Chazy spacetime the answer is positive. If one approaches the origin $\rho^2+z^2=0$ properly (along the z-axes) one can then pass through the ring-like singularity and enter Minkowski space which can be smoothly connected to the Curzon--Chazy spacetime. This was shown by Morgan and Szekeresz \cite{Morgan:1973va} and naively it seems that it should also be possible in the SCC spacetime whose curvature invariants are everywhere finite. However the finiteness of curvature scalars does not mean an absence of a curvature singularity. If a component of the Riemann tensor with respect to a parallel-propagated orthonormal frame diverges we can also speak of a curvature singularity which is then called non-scalar. This is well described in \cite{nonsingular}.  \par 
Recalling the fact that the radial timelike geodesics are incomplete we may find a free-falling frame  $\lbrace e_\nu^\mu \rbrace_{\nu=0...3} $ where $e_0^\mu $ is the four-velocity of the radial geodesic \eqref{fourvelocity}. The orthonormal frame is then

\begin{equation}\label{tetradgeod}
 \begin{split}
e_0^\mu &=(E e^{\frac{2m}{r}},-\sqrt{E^2-e^{\frac{-2m}{r}}},0,0), \hspace{15pt} e_1^\mu =(e^\frac{m}{r}\sqrt {{E}^{2}{{\rm e}^{\,{\frac {2m}{r}}}}-1}
,-E,0,0)\\
e_2^\mu &=(0,0,{\frac {1}{r}{{\rm e}^{-{\frac {m}{r}}}}},0), \hspace{15pt} e_3^\mu =(0,0,0,{\frac {1}{r \sin(\theta)}{{\rm e}^{-{\frac {m}{r}}}}}).
 \end{split}
\end{equation} 
The frame is indeed parallel-propagated along the geodesic as it satisfies the equation $e_0^\alpha \nabla_\alpha e_\nu^\mu=0$. We can then compute the frame components of the Riemann tensor and find out that many of them indeed diverge in the limit $r\rightarrow 0$. In particular two components of the electric part of the Riemann tensor tensor have the form

\begin{equation}\label{riemann}
\tensor{R}{_0^2_0_2}=\tensor{R}{_0^3_0_3}=-{\frac {m}{{r}^{4}} \left( m {E}^{2}-r{{\rm e}^{-\,{\frac {2m}{r}}}}
 \right) }.
\end{equation} 
These components play role in the geodesic deviation equation which can be written in our frame as

\begin{equation}\label{geoddev}
\frac{\mathrm{d^2}n^i}{\mathrm{d}\tau^2}=-\tensor{R}{_0^i_0_j} n^j.
\end{equation} 

We can thus conclude that the free falling observer would experience infinite tidal forces in the directions tangent to the sphere $r=const.$ when approaching $r=0$. This is consistent with the behaviour of the areal radius \eqref{radius} which grows to the infinity in the limit as well. When examining the situation more closely we found out that the frame components of the Weyl tensor are perfectly finite. It is therefore the Ricci part of the curvature which is singular. In the limit $r\rightarrow 0$ all the nonzero components of the Ricci tensor tend to 

\begin{equation}\label{Ricci}
\tensor{R}{_0_0} , \tensor{R}{_1_0}, \tensor{R}{_1_1} \rightarrow -\,{\frac {{2E}^{2}{m}^{2}}{{r}^{4}}} .
\end{equation} 

This is not very surprising as the Ricci tensor is given by the matter filling the spacetime and in our case  the scalar field diverges in $r=0$. For illustration we can compute the components of the  Ricci tensor in a different frame. We know  that the distance along a spatial geodesic located at $t=const.$ is infinite (equation \eqref{properdistance}). Avoiding the geodesic incompleteness here should yield finite curvature components. And it is indeed so. If we use

\begin{equation}\label{tetradgeod2}
\hat{e}_0^\mu=( e^{\frac{m}{r}},0,0,0), \hspace{15pt} \hat{e}_1^\mu=(0,e^{-\frac{m}{r}}
,0,0)
\end{equation} 
together with $e_2$ and  $e_3$  from \eqref{tetradgeod} where $\hat{e}_1$ is the tangent vector we obtain the only nonzero Ricci component
\begin{equation}\label{Ricci2}
\tensor{\hat{R}}{_1_1} =-\,{\frac {2{m}^{2}}{{r}^{4}}{{\rm e}^{-\,{\frac {2m}{r}}}}}
\end{equation} 
which is finite. The parallel transport equations are again satisfied for this frame ($\hat{e}_1^\alpha \nabla_\alpha \hat{e}_\nu^\mu=0$).

The hypersurface $r=0$ is then a Ricci curvature singularity which is encountered in some frames  while in others the curvature is regular.  We can even find coordinates  in which the Ricci tensor is finite. For $x=e^{\frac{m}{r}}$  the scalar field is $\Phi=\mp\ln(x)$ and $\tensor{R}{_1_1}\sim \frac{1}{x^2}$ which is zero for $x\rightarrow \infty$. The scalar field is however singular no matter which frame we use and therefore it would be surprising if the hypersurface $r=0$ were regular and we could extend the spacetime beyond this boundary. {In fact, even without considering the scalar field, the divergent components of the Riemann curvature tensor (as seen in \eqref{riemann}) in the parallel‐propagated frame \eqref{tetradgeod} rule out the existence of any extensions (see \cite{Clarke}). }

\section{A note on $m<0$ case}
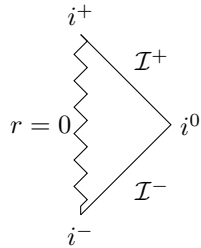
\begin{figure}[ht]\centering
\begin{tikzpicture}[scale=0.3]
\node (I)    at ( 4,0)   {};
\path % Four conners of the right diamond (no labels this time)
   (I) +(90:4)  coordinate[label=90:$i^+$] (Itop)
       +(-90:4) coordinate[label=-90:$i^-$] (Ibot)
       +(180:4) coordinate (Ileft)
       +(0:4)   coordinate[label=0:$i^0$] (Iright)
       ;
% No text this time in the next diagram
\draw  (Itop) --
       node[midway, above right]    {$\mathcal{I}^+$}  
 (Iright) -- 
 node[midway, below right]    {$\mathcal{I}^-$}  
 (Ibot); 
\draw[decorate,decoration=zigzag] (Itop) -- (Ibot)
 node[midway,left] {$r=0$};    
\end{tikzpicture}
\caption{Penrose conformal diagram for $m<0$ case.\label{nakedsingularity}}
\end{figure}

Finally, let us briefly comment on the case of $m<0$ which we have disregarded so far. Examining \eqref{Kretschmann} we immediately see that curvature singularity at $r=0$ is now a simple scalar one since Kretschmann scalar diverges. Additionally, the character of the singularity ($r=0$ hypersurface) changed from null to timelike and it has repulsive nature since $m$ characterizes asymptotic mass of the spacetime. The singularity is practically inaccessible for timelike observers because it is surrounded by infinitely high potential barrier and therefore only observers accelerated to attain $E=\infty$ could reach $r=0$. The null observers can reach the singularity without any problem. Also the radial distance to it $d(r_0,0)$ is finite as well as the coordinate time $t$ it takes to reach $r=0$, in fact these quantities are finite for all radial geodesics. The wormhole structure is absent completely as there is no throat (the expansions of relevant congruences are nonzero everywhere). Overall, the spacetime has a conformal structure of a simple timelike naked singularity (see figure \ref{nakedsingularity}).

\section{The Maxwell-scalar solution}

When considering the metrics of the form \eqref{Gibbons} one can generalize the method of generating new solutions to also include the electromagnetic field. This was done for example in  \cite{Maeda}.  The action \eqref{action} is then extended by including the Lagrangian density of the Maxwell field 
\begin{equation}\label{EMLagrangian}
 \mathcal{L}_{\mathrm{EM}} = -\frac{1}{4} F_{\mu\nu}F^{\mu\nu}
\end{equation}
which means adding the stress-energy tensor  
\begin{equation}\label{EMTmunu}
T^{(\mathrm{EM})}_{\mu\nu}= F_{\mu\rho}\tensor{F}{_\nu^\rho} + \mathcal{L}_{\mathrm{EM}}\, g_{\mu\nu} 
\end{equation}
to the right-hand side of the Einstein equations. Based on \cite{Maeda}, the solution generated from our SCC spacetime by including the electromagnetic field has the following form

\begin{equation}\label{EMmetric}
\begin{split}
\displaystyle\mathrm{d}s^2 &=-\frac{1}{\sinh^2(\frac{m}{r})}\mathrm{d}t^2+\sinh^2\Big(\frac{m}{r}\Big)(\mathrm{d}r^2+r^2 \mathrm{d}\Omega^2), \\
\Phi &=\pm\frac{m}{r}, \qquad F_{tr}=\pm\frac{m}{r^2\sinh^2(\frac{m}{r})} .
\end{split}
\end{equation}
The relation to the SCC spacetime becomes clear when looking at the the dependence of the metric on the radial coordinate $r$. For $r\rightarrow 0$ the dominant source is the scalar field while the Maxwell field vanishes in the limit. Thus in the neighborhood of $r=0$ the metric \eqref{EMmetric} reduces to the SCC metric \eqref{scc}. On the other hand as $r\rightarrow \infty$ the strength of the electromagnetic field grows to $F_{tr} = \pm\frac{1}{m}$. This in turn has effect on  energy conditions as a timelike observer would measure a positive energy density while the null energy condition remains unaffected by $T^{(\mathrm{EM})}_{\mu\nu}$.\par
\begin{figure}[h!]\centering
\begin{tikzpicture}[scale=0.4]
\node (I)    at ( 4,0)   {};
\path % Four conners of the right diamond (no labels this time)
   (I) +(90:4)  coordinate[label=90:$i^+$] (Itop)
       +(-90:4) coordinate[label=-90:$i^-$] (Ibot)
       +(180:4) coordinate (Ileft)      ;
% No text this time in the next diagram
\draw (Itop) 
 (Ibot);
    
\draw[decorate,decoration=zigzag] (Itop) -- (Ileft)
      node[midway, above, outer sep=5mm] {$r=0$};

\draw[decorate,decoration=zigzag] (Ileft) -- (Ibot)
      node[midway, below, outer sep=5mm] {$r=0$};       
\draw (Itop) -- (Ibot)
 node[midway,right] {$r=+\infty $};    
\end{tikzpicture}
\caption{The conformal diagram of the SCC spacetime with electromagnetic field.} \label{EMPenrose}
\end{figure}
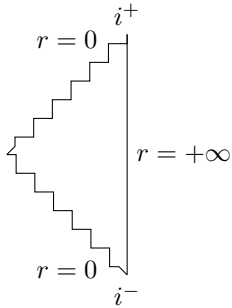

The areal radius $R_A(r)$ is a decreasing function of $r$ which means that no wormhole is present in this case (The expansions of null radial geodesics do not change their signs as well.). Since the electromagnetic field does not vanish at the infinity it is clear that the spacetime is not asymptotically flat which is also apparent from its conformal diagram (figure \ref{EMPenrose}) where the infinity ($\vert t\vert<\infty, r =\infty$) is a timelike hypersurface. On the other hand the left part of the diagram including the non-scalar singularity is the same as in the SCC spacetime. 

The timelike character of the infinity looks similar to an asymptotically anti-de Sitter spacetime that is however not the case here as the scalar curvature vanishes at infinity. When looking at the metric \eqref{EMmetric}  at $r=\infty$   we can see that the areal radius reaches its minimal value $m$ while the other metric components do not converge to a finite nonzero value. It is important to stress however  that nothing suggests that there is a spacetime singularity at  $r=\infty$. Firstly none of the timelike geodesics can reach it due to the presence of an infinite potential barrier. As for the spacelike and null geodesics, all of them reach it at an infinite  value of their affine parameter, in addition all  components of the Riemann tensor remain finite in the limit.\par
Since the spacetime is not asymptotically flat (or even simple) the concept of total mass-energy of the spacetime is  ambiguous. One can use the Brown--York energy  and evaluate the integral \eqref{BY} at infinity. As in the SCC case however it is necessary to use the areal radius as a radial coordinate to get a meaningful result (the coordinate $\rho$ in \eqref{euclidr}). One then gets the expected result $M_{BY}=m$. If one uses the Misner--Sharp energy (which is a coordinate-independent quantity), one arrives at $M_{MS}=\frac{m}{2}$. The difference between the two masses is not that surprising since they coincide for a general spherically symmetric spacetime only at such $\rho_0$  for which $g_{\rho\rho} (\rho_0)=1$. This can be seen from the general forms of both masses (see for example \cite{BYgeodesics}) and it is the reason why they give the same result for a metric converging to Minkowski one. The Komar mass \eqref{Komar} is of no use here as it produces an infinite result in this case.

\section{Conclusion}
We have seen that the solution we examined is behaving both as a wormhole and a time machine. However, neither of these characterisations are fully attained by our solution. Although we have a wormhole throat satisfying all the standard conditions it connects asymptotically flat region to a region whose border is a non-scalar singularity of null character. This strange region enables travelling to future timelike infinity in finite time thus essentially being a one-directional time machine. Overall picture of the geometry was provided by analysing conformal structure.

The parameter $m$ characterising the solution can be interpreted as a mass of the spacetime as confirmed by using several definitions. The Misner--Sharp energy turned out to be a good tool for localization of the energy distribution in the spacetime. These results were compared with Katz, Lynden-Bell and Bi\v{c}\'{a}k approach \cite{Katz_2006}. 

A brief study of the case $m<0$ revealed that it results in a spacetime that looks conformally as a standard timelike naked singularity. However, this singularity is surrounded by infinite potential barrier making it inaccessible to observers. Finally, a generalization including electromagnetic field was obtained by applying a generating technique leading to a spacetime with radically different conformal structure and without wormhole throat.

The non-scalar singularity is rather undesirable feature of our spacetime but one expects that quantum gravity effects would remove it as is the case for the timelike naked singularity sourced by scalar field \cite{Tahamtan-Annals:2020}. Another interesting aspect would be a potential generalization including rotation. {Considering the one-directional time machine character of our solution it might be interesting to analyze if the extra positive time gained by passing in and out of the wormhole might help in protecting causality in systems composed from several such wormholes.}

\section*{Acknowledgements}

The research was supported by the Czech Science Foundation Grant No. 22-14791S and the Charles University project GAUK No. 906419.

\section*{References}
\bibliographystyle{iopart-num}
\bibliography{sample}
\end{document}